# Modeling the System-Level Reliability towards a Convergence of Communication, Computing and Control

Bin Han[*,1] and Hans D. Schotten[1,2]

[1]*Division of Wireless Communications and Radio Positioning (WICON), University of Kaiserslautern, Germany*
[2]*Research Department Intelligent Networks, German Research Center for Artificial Intelligence (DFKI), Germany*

**Abstract**

Enabled and driven by modern advances in wireless telecommunication and artificial intelligence, the convergence of communication, computing, and control is becoming inevitable in future industrial applications. Analytical and optimizing frameworks, however, are not yet readily developed for this new technical trend. In this work we discuss the necessity and typical scenarios of this convergence, and propose a new approach to model the system-level reliability across all involved domains.

## 1. Background

Research interest on the communication-computation-control convergence (Co4) has been raised since almost two decades. Dating back to 2002, *Mitter* proposed in [1] that in an automated system, not only the sensors, but also the communication modules and the computational algorithms, shall be considered as part of the controlling system, since they must all associate with a specified controlling scenario and serve for a control goal. One year later, for the sake of system-level performance optimization, *Graham* and *Kumar* also called for a holistic theory that connects the fields of control, communication and computation. They also implemented a testbed of wireless remote vehicle controlling [2]. However, as the demand for such a system theoretic model across different domains was not urgent at that time, little follow-up effort had been made until a few years ago.

The recent revival of interest in this area was driven by the arising demand for future B5G and 6G networks. In context of the next generation mobile system, the ever-more complex controlling tasks, such as autonomous driving, context-aware CPS and autonomous manufacturing, are introducing increasingly complex computing problems. Such tasks will probably rely on machine learning and artificial intelligence solutions, which are commonly computationally challenging. On the other hand, sensors and actuators are not only becoming more and more compact to enable dense deployment and ubiquitous integration into *everything*, but also consuming lower and lower energy for better sustainability. The limits in size and power generally prohibits local execution of the complex computing algorithms, so that edge/cloud solutions have to become liable, where the communication module plays an indispensable role.

This new paradigm for controlling systems proposes emerging technical challenges. The modern industrial applications are calling for high deployment density (e.g. IoE) or/and ultra-low latency (autonomous manufacturing), which likely leads to unreliable communication links. Meanwhile, the industrial scenarios generally have ultra-high requirement of dependability, which can be very likely violated by the communication failure. To make it worse, there has been yet no quantitative model that describes the impact of communication link reliability on the controlling system reliability. One key reason is the fundamental difference in the definitions to reliability in the two communities. More specifically, the reliability of a communication link is usually evaluated by its error rate / packet loss rate / etc. under a given latency constraint. The reliability of a controlling system, on the other hand, is commonly described in terms of stability margins, failure rate, mean time between failures (MTBF), mean time till failure (MTTF), etc. The problem can become even more complicated, when the impact of computing module is introduced as well. Towards a Co4, the need for a cross-layer reliability model can no more be evaded.

## 2. Examples of Co4 Scenarios

Though no generic Co4 reliability model is available yet, there have been various scenarios investigated, where Co4 is at least partially concerned. In this section, we name and briefly discuss some examples.

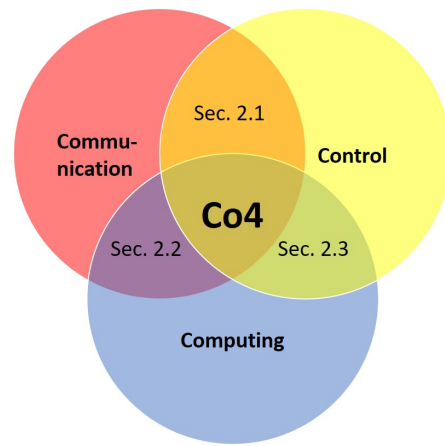

**Figure 1.** Sub-domains of CoCoCo discussed in this section

### 2.1. Communication-Control Codesign (CoCoCo)

As one of the most typical and important Co4 scenario, existing studies in CoCoCo mostly focus on the codesign and joint optimization of sensing and communication. For example, the optimal status update in networked systems has been widely studied based on the concept of age of information (AoI) [3].

Generally, the AoI of a wireless networked system is jointly determined by the updating at information source and the communication channel between source and sink.

---
[*]E-Mail: bin.han@eit.uni-kl.de

On the one hand, a lower update rate is pushing the AoI higher from the perspective of information source, regardless of the communication quality. On the other hand, a higher update rate implies more messages to be transmitted over the wireless channels, which will reduce the communication reliability and therewith cancel out the benefit brought by the fresher data at source.

More specifically, in the Shannon regime where error-free transmission can be guaranteed for each message by occupying the channel for sufficiently long time, the competition among numerous messages for limited radio resource can lead to channel congestion, which is usually reflected in a high queuing delay. This phenomenon has been observed since the emergence of AoI concept [4], and deeply investigated in various practical application scenarios such as [5]. In the finite-blocklength (FBL) regime, to the contrary, we can shorten the channel occupation time of each message to resolve the MAC congestion, but at a price of higher packet error rate (PER), which causes numerous packet losses and re-transmissions that increases the AoI. This phenomenon has been well studied in our previous work upon a blind-scheduled sensor network scenario [6].

In the end, the AoI commonly becomes a convex function of the update rate. While the state estimation error in controlling loop is a monotonically increasing function of the AoI [7], [8], the update policy can be optimized to minimize the AoI, and therewith also the system error.

A recent trend in research in this field is to extend AoI into the so-called effective AoI, a.k.a. age of incorrect information (AoII). While AoI takes a pure networking perspective in which the communication module does not need to understand the information, AoII is only applicable for semantic communications where the communication module needs to know the utility of every piece of information it transmits. Thus, it is more close to the future Co4 paradigm. The impact of this transition from AoI to AoII can be typically observed from the comparison between [9] and [10]: the earlier proposes that a $D/M/1$ queuing system with periodic sensing is superior over an $M/M/1$ one with event-triggered sensing under same conditions, since it delivers significantly lower AoI; the latter shows that, however, event-triggered sensing works better than periodic sensing by means of lowering the AoII.

### 2.2. Communication-Computing Co-Design

#### 2.2.1. Optimal Task Offloading

This has been one of the most studied topics in cloud computing and MEC domain, where the user equipment (UE) can choose between local computation (high computation latency, high computing energy consumption) and cloud/MEC offloading (low computation latency, no computing energy consumption, channel-state-dependent latency and energy consumption for communication, sometimes with random server queuing latency). The most commonly considered criteria are latency and energy efficiency [11], upon certain scenario and requirement, other criteria such as security can also be taken into account [12].

#### 2.2.2. Optimal Pre-Prosessing

In this scenario it considers an optional pre-processing of information (e.g. classification, compression, feature extraction, etc.) before the transmission can reduce the time and power needed for the communication. However, the processing itself takes time and power. So that an optimal pre-processing policy that maximizes the overall performance is of interest. A pioneering study on the simplified model of on-off block fading channel regarding minimal long-term average AoI is provided in [13].

### 2.3. Computing-Control Co-Design

#### 2.3.1. Sharing Computing Resource among Controlling Tasks

The optimal scheduling of limited computing resource (e.g. CPU time) among multiple independent controlling tasks, known as the control-scheduling co-design, has been one of the most important research topic in the field of networked control systems [14].

#### 2.3.2. Optimal Observation and Estimation

Consider a control algorithm that makes decisions based on its estimation to an unknown/hidden system status, which must be learned (computing) from the observations (sensing). The longer the learning process takes, the lower the estimation error. However, it lowers the controlling rate and reduces the system reliability. This trade-off between latency and estimation error plays an important rule in the design of robust networked control system [15]. In a recent study [16], we have also investigated this problem in scope of MEC task offloading regarding queue management.

## 3. Full-Stack Modeling of System-Level Reliability

### 3.1. Inverted Pendulum: A Case Study

To demonstrate the necessity and complexity of system-level reliability modeling, we carry out a numerical simulation of inverted pendulum system remotely controlled over wireless channels. The system is illustrated in Fig. 2, where the objective is to control the desired position $q(t)$ of the cart with an inverted pendulum (see Fig. 2a), in order to track a reference signal $r(t)$ while stabilizing its pendulum angle $\theta(t)$. The initial position is set to $q(0) = 0$, and the $r(t)$ is a rectangular pulse with a period of 20 s and duty ratio of 40%, ranging from 0 m to 5 m.

We consider an adjustable control interval $T_\text{ctr}$ of the controller, which is also to the sampling interval of the sensors, which are integrated in the cart to measure the state vector $\mathbf{x} = [q, \theta, \dot{q}, \dot{\theta}]^\text{T}$. To maximize the radio resource utilization over wireless channels, we also consider the transmission frame length in both uplink (UL) and downlink (DL) to be $T_\text{ctr}$, as detailed in Fig. 2b.

For the communication module, we consider short-coded messages to be sent with $d = 64$ bit payload in both UL and DL. It uses BPSK modulation transmits over a carrier bandwidth $B = 1$ kHz with $r_\text{s} = 1$ kSPS symbol rate, where both UL and DL channels are AWGN channels with the SNR of $\gamma = 0$ dB. The PER can be then approximately

estimated according to [17]:

$$\varepsilon \approx Q\left(\sqrt{\frac{r_s T_{\text{ctr}}}{V}}\left(\frac{C}{B} - \frac{d}{r_s T_{\text{ctr}}}\ln 2\right)\right), \quad (1)$$

where $C = B\log_2(1+\gamma)$ is Shannon's channel capacity and $V = 1 - \frac{1}{(1+\gamma)^2}$ the dispersion of AWGN channels.

To evaluate the control performance, we define a quadratic cost

$$c = \frac{\sum_{k=0}^{K}\left\{0.5\theta^2(t+kT_{\text{ctr}}) + 0.5\left[q(t+kT_{\text{ctr}}) - r(t+kT_{\text{ctr}})\right]^2\right\}}{K+1} \quad (2)$$

where $K = \left\lfloor\frac{6000}{T_{\text{ctr}}}\right\rfloor$ is the number of samples over an simulation episode of $6000\,\text{s}$. We tested the control cost $c$ and the PER $\varepsilon$ for different values of $T_{\text{ctr}}$, with 50 episodes of independent Monte-Carlo test per specification, the results are shown in Fig. 3. We can observe a convexity where the optimum occurs around $T_{\text{ctr}} \approx 9\,\text{ms}$ (see zoomed part in Fig. 3b). When further decreasing the control interval, the loss caused by a significant PER over 1% will overwhelm the benefit of timely update. In contrast, when assuming ideal communication channel, the control cost is monotonic about $T_{\text{ctr}}$. This phenomenon reveals a necessity of system-level reliability modeling.

### 3.2. Error Propagation Model
#### 3.2.1. Error Classification

In Co4 approaches, "errors" defined in different domains must be clearly distinguished from each other: there are

1. Value errors, such as sensing error, computation error, source coding noise, quantization noise, system disturbance, etc. Each of these errors can be generally considered as an additive disturbance at some certain stage of the system, and directly leads to a bias or divergence of the system from its desired state;

2. Transmission errors. Modern digital communication systems are equipped with sound error-control mechanisms such as error detection code, forward error correction code, and automatic repeat request. It can be sufficiently trusted that decoding errors in such systems will not be converted to value errors, but only cause retransmissions or packet losses. In both cases, it leads to an increased inter-arrival time of messages, either at the controller (i.e. feedback latency) or at the actuator (i.e. control latency).

3. System errors, which are observed on the application layer and cause failure or incapability of executing the desired task.

#### 3.2.2. Modeling the Error Propagation: Methodologies

A key challenge here is to figure out, how value errors and transmission errors impact the system errors. Generally, two kinds of approaches can be taken. Analytically, tools have been developed in the fields of stochastic control and networked control system to model the control performance regarding random feedback latency. These models

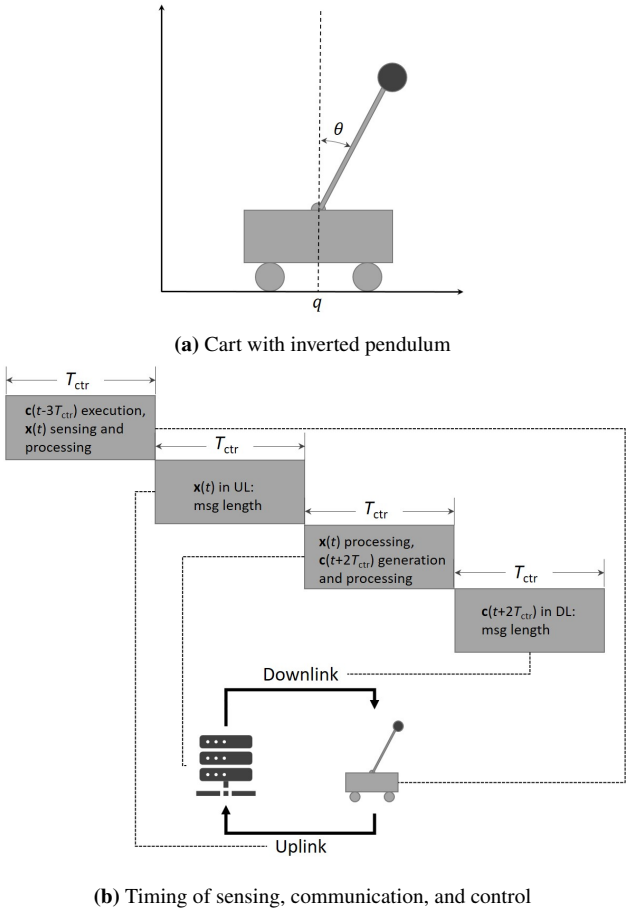

(a) Cart with inverted pendulum

(b) Timing of sensing, communication, and control

Figure 2. The simplified inverted pendulum simulation

shall be merged with communication models such as 1 and computing models, to provide deep insights on the impact of value and transmission errors on system errors. Alternatively, exhaustive numerical simulation campaigns can be conducted to build empirical models.

### 3.3. The New Approach to Model Reliability: A System Engineering Perspective

The traditional approaches, as illustrated in Fig. 4, generally consider controlling logic as endogenous and embedded in the application, which proposes a specific QoS requirement on the communication module. The parameters of the controlling logic itself, such as sensing rate and controlling interval, however, are usually selected with unnecessarily rich margins, since their design is decoupled from the communication and computing modules and cannot rely on them. Thus, even on the radical horizon of optimization in the communication/computation modules, there will still be a lot of potential wasted by the conservative controlling design.

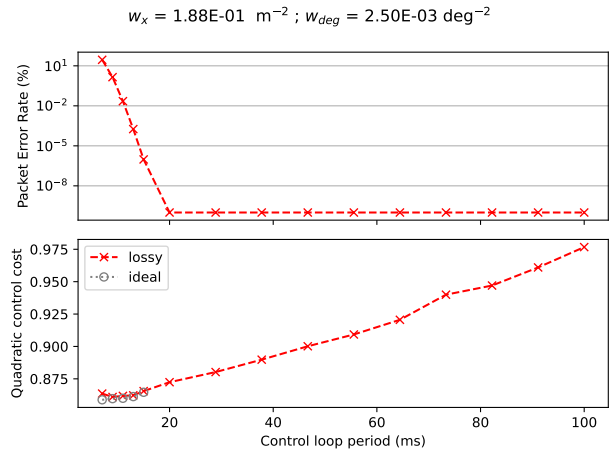

**(a)** The PER and control cost as functions of control interval

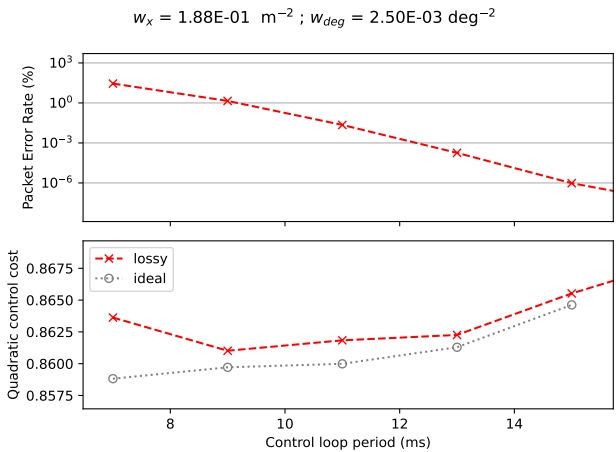

**(b)** Zoom-in

**Figure 3.** The simulation results

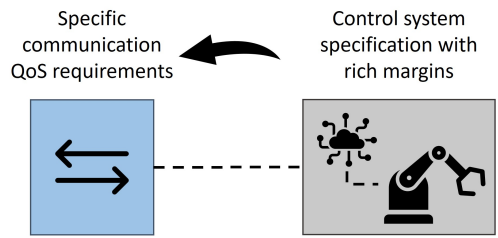

**Figure 4.** Traditional methodology of networked control system design

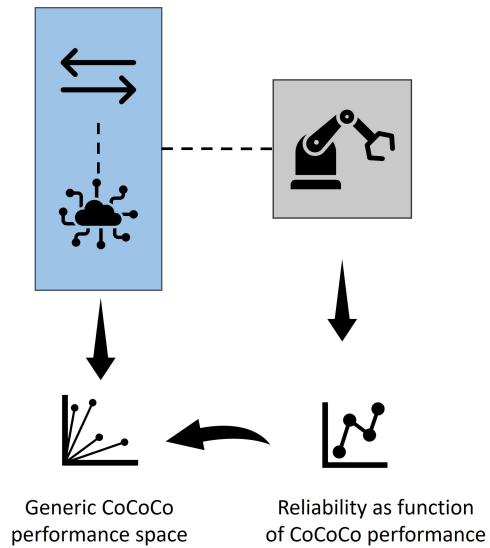

**Figure 5.** Proposed novel approach of networked control system design

In contrast, we propose to take a novel view that consider the CoCoCo part as a blackbox, which is paired with an abstracted application module regardless its physical implementation, as illustrated in Fig. 5. On the one hand, every CoCoCo system has its feasible region in a generic "performance space", which is the set of combinations of

AoI/AoII and generalized error vector. A CoCoCo optimization can expand this feasible region to its utter best. On the other hand, every specific application can be characterized with a performance-reliability model, which can be set up through satisfactory tests. Generally, the feasibility region shrinks as the required minimal reliability level rises. In this way, all modules in the Co4 system are able to be jointly optimized. Since the traditional approach takes only a static point or subset from the feasible region, we know it is outperformed by the proposed approach.

## 4. Conclusion

In this paper, we have visited the problem of system-level reliability modeling regarding the technical trend of communication-computing-control convergence. We have reviewed the types of sub-Co4 problems that have been studied, and discussed the full-stack modeling of system-level reliability. More specifically, we have demonstrated the necessity of such model through a simplified simulation of inverted pendulum, briefly investigated the error propagation problem in Co4, and proposed a new approach from a system engineering perspective.